\documentstyle[aas2pp4]{article}

\lefthead{Bowyer and Bergh\"ofer}
\righthead{Inverse Compton Scattering in the Coma Cluster}

\begin{document}

\title{Inverse Compton Scattering as the Source of Diffuse EUV Emission in the
Coma Cluster of Galaxies}

\author{Stuart Bowyer and Thomas W. Bergh\"ofer}

\affil{Space Sciences Laboratory, University of California, Berkeley, 
        CA 94720-7450, USA}

\begin{abstract}
We have examined the hypothesis that the majority of the diffuse EUV flux in 
the Coma cluster is due to inverse Compton scattering of low energy cosmic ray 
electrons ($0.16 < \epsilon < 0.31$\,GeV) against the 3$^{\circ}$\,K black-body
background. We present data on the two-dimensional spatial distribution of the
EUV flux and show that these data provide strong support for a non-thermal 
origin for the EUV flux. However, we show that this emission cannot be 
produced by an extrapolation to lower energies of the observed synchrotron
radio emitting electrons 
and an additional component of low energy cosmic ray  electrons is required. 
\end{abstract}

\section{Introduction}

Diffuse EUV emission has been detected in five clusters of galaxies:
Virgo (Lieu et al., 1996a), Coma (Lieu et al., 1996b), Abell 1795 
(Mittaz et al., 1997), and Abell 2199 and Abell 4038 (Bowyer et al., 1997).
These clusters were detected with a statistical significance varying from 8 to
50 standard deviations. The diameter of the diffuse emission in these clusters
ranges from 20$\arcmin$ to 40$\arcmin$.

Some diffuse EUV emission in clusters of galaxies would be expected from the 
well-studied X-ray cluster emission. However, in all 
cases examined to date, the EUV emission is far greater than the expected 
emission from the X-ray-emitting gas. Marginal signatures of this ``soft 
excess'' are sometimes present in the lowest energy resolution band
of the ROSAT PSPC, where they produce less than a 20\% enhancement
over the emission expected from the X-ray cluster gas.
In contrast, the excesses found with EUVE range from 70\% to 600\%
above that expected from the X-ray gas.
A variety of instrumental and Galactic interstellar medium absorption effects
have been suggested as alternative explanations for the EUVE data, but these 
have all been found wanting (For a discussion of these issues, see Bowyer et 
al., 1997).

In the original reports of diffuse EUV cluster emission, the data were 
interpreted in terms of additional thermal gas components in the clusters.
In these analyses, the known X-ray emission from the hot cluster gas
was first fitted to the EUV data, and the excess EUV emission was computed.
This excess was then fitted by additional components of thermal gas.
Because of the low energy of the EUV emission, much lower temperature
thermal gases ($\sim 10^6$\,K) are required. The concept that additional 
components of lower temperature gas are present in these clusters has not 
received wide support, primarily because gas at these temperatures is near the
peak of the radiative cooling curve and hence cools rapidly, requiring a
substantial energy input to sustain the gas at these temperatures.
In addition, it is difficult to understand how different components of gas at
grossly different temperatures could retain their separate identity. For
example, the Coma cluster has been shown to be formed by merging of distinct
subunits, in both X-ray (White et al., 1993), and optical (Colless and Dunn,
1996) studies. This produces variations less than a factor of 3 in
temperatures of the X-ray emitting gas (Honda et al. 1996). Deiss and Just
(1996) have shown in a general analysis, and with a specific application to
the Coma cluster, that turbulent mixing time scales are only a few
$10^9$\,years, which argues against the co-existence of major quantities of
gas at two vastly different temperatures. However, Cen et al. (1995) have
argued that a warm ($\simeq 10^6$\,K) thermal gas is widely distributed
throughout the universe, as a direct product of the growth of structure
leading, eventually, to clusters of galaxies. In their scenario, the energy
required to sustain the warm gas is provided by gravitation.

Hwang (1997) has examined the hypothesis that the source of the diffuse
EUV flux is inverse Compton scattering by electrons that are
a low energy extrapolation of electrons producing the observed synchrotron 
emission; these electrons are scattered  against the 3$^{\circ}$\,K black-body 
background radiation. The magnetic field he derived for the cluster was 0.2 to
0.4\,$\mu$G which is consistent with the range of estimates for the cluster 
field. In this work he only considered the constraints imposed by the total 
EUV flux. 

En{\ss}lin and Biermann (1998) also considered this mechanism as the source of 
the EUV flux in the Coma cluster. They assumed that the relativistic energy 
density of the synchrotron emitting electrons scale radially with the same 
profile as the X-ray producing gas. This assumption can be questioned since 
the non-thermal relativistic electrons may well be independent of the thermal 
X-ray gas. These authors cite the best support for this assumption is given by
the data in Figure\ 3 of Deiss et al. (1997) which compares the X-ray and radio
radial emission profiles. Unfortunately, the results in this figure are 
incorrect as has been confirmed by Deiss (private communication).
These authors find a magnetic field of 1.2\,$\mu$G is required given their
assumptions; this field is also consistent with the range of estimates for the 
cluster. 

Sarazin and Lieu (1998) have explored the possibility that
EUV radiation in clusters of galaxies could be produced by inverse Compton
scattering by a population of very low energy cosmic ray electrons.
They showed that the one dimensional EUV spatial profile for the cluster A1795,
a radio quiet cluster, was consistent with this hypothesis. A potential 
problem with this hypothesis as a universal explanation for the EUV emission 
in clusters of galaxies is that the electrons they proposed have an energy 
density and pressure which are 1 to 10\% of that of the thermal gas in 
clusters. If one includes the pressure of cosmic ray ions to the pressure of 
the electrons proposed by Sarazin and Lieu, using the ratio expected on 
theoretical grounds (Bell 1978) and measured at Earth orbit (see. e.g., Weber 
1983), the total cosmic ray pressure is substantially larger than that of the 
X-ray emitting gas. 

In this work we reconsider the hypothesis that the EUV emission in the Coma
cluster is the result of inverse Compton emission. We first consider the 
constrains imposed by the total EUV flux. We review the existing radio data 
and obtain a different spectral index than that employed by Hwang (1997) and
En{\ss}lin and Biermann (1998), which we argue is more appropriate. We then 
derive results which are generally consistent with the inverse Compton 
hypothesis. We then consider the two dimensional spatial distribution of the 
EUV flux; this data provides substantial support for a non-thermal origin for 
this flux. We find that the spatial distribution of the magnetic field 
required by En{\ss}lin and Biermann to produce the EUV emission profile is 
unrealistic. We show that the difference in spatial extent between the EUV and
radio halos cannot be explained using an electron distribution which is an 
extrapolation of the known synchrotron emitting electrons and that an 
additional population of low energy cosmic ray electrons are required to 
explain these data.

\section{The Total Inverse Compton Emission}

Assuming that relativistic electrons produce the radio emission by synchrotron
radiation, the observed power law radio spectrum can be used to derive the 
number spectrum of the electrons in the cluster, which is normally
characterized by a power law spectrum of the form:
\begin{equation}
N(\gamma) = N_0 \cdot E^{-\gamma} .
\end{equation}
Following Pacholczyk (1970), the synchrotron emissivity of such an ensemble of
relativistic electrons is 
\begin{equation}
\epsilon_{\nu} = c_1(\gamma)~N_0~B^{(\gamma+1)/2}~\left(\frac{\nu}{2 c_2}\right)^{(1-\gamma)/2} ,
\end{equation}
where $\nu$ is the radio frequency, $B$ the magnetic field strength 
perpendicular to the line of sight, $c_1$ is a function of the spectral index 
of the electron spectrum $\gamma$ (see Pacholczyk 1970), and $c_2$ a constant.
The observed radio flux is related to the emissivity by 
$S_{\nu} = \epsilon_{\nu} / D^2$, where $D$ is the cluster distance.
(In this work we assume H = 50\,km\,s$^{-1}$Mpc$^{-1}$) For the 
same population of electrons the total inverse Compton flux (cf. Blumenthal and
Gould 1970, Ginzburg and Syrovatskii 1964) is
\begin{eqnarray}
F_{ic} & = & \frac{1}{4 \pi D^2} \frac{8 \pi^2 e^4}{h^3 m_e^2 c^6} f_1(\gamma)
N_0 (m_e c^2)^{-\gamma+1} \\ 
& & \times (k_B T)^{(\gamma+5)/2} E_{ic}^{-(\gamma-1)/2} ,\nonumber
\end{eqnarray}
where $E_{ic}$ is the energy of the comptonized photons, $T$ the temperature 
of the microwave background, $k_B$ Boltzman's constant, and $f_1$ a function 
of $\gamma$. The observed radio flux can be related to the inverse Compton 
flux by inserting $N_0$ from eq. (2) which leads to the following relation
\begin{equation}
F_{ic} \propto f(\alpha) F_{sync} B^{-(\alpha + 1)} (k_B T)^{\alpha+3} E_{ic}^{-\alpha} ,
\end{equation}
where $\alpha = (\gamma -1)/2$, $S$ is the radio flux,
and $f$ a function of $\alpha$.

We first review the observational data on the intensity and spectral index of
the synchrotron emission in the Coma cluster. We note that observations of 
synchrotron emission in clusters are quite difficult. Major problems include 
the difficulty of detecting low surface brightness diffuse radiation in regions
containing blended point sources of emission, and a variety of technical 
difficulties, including problems in obtaining adequate integration volumes
and correct intensity offsets. A number of authors have provided summaries
of quality observations of synchrotron emission in the cluster, measured
at frequencies ranging from 10.3\,MHz to 4.85\,GHz.

We are especially interested in lower frequency radio observations, since
the electrons producing this emission are closer to the energies which will 
produce the inverse Compton EUV flux. With a magnetic field in the range of 
$\approx$0.1 to $7 \mu$G (as discussed in the following), electrons with 
energies in the range 0.16 to 0.31\,GeV will produce EUV inverse Compton 
emission. The corresponding synchrotron emission will range from 4 to 
15\,MHz for a $0.1 \mu$G field and 0.4 to 0.15\,MHz for a $1 \mu$G field.

The lowest reported detection of radio flux from the Coma cluster was by 
Bridle and Purton (1968), who carried out transient scans of 124 radio sources
at 10.3\,MHz using the Dominion Radio Astrophysical Observatory
Low Frequency Array. Ionospheric 
effects were substantial at this frequency. The major thrust of this work was 
simply to detect the total low-frequency flux from sources in the Revised 
Third Cambridge Catalogue of Radio Sources, and detailed studies of individual
sources were not carried out. The ratio of measured flux to expected flux for 
these sources varied by a factor of ten; the extent to which this was an 
observational effect or was due to true low frequency cutoffs in the sources 
is not known. We believe this measurement to be sufficiently uncertain that we 
will not consider it further.

The next lowest frequency observation was made by Henning (1989).
She used the Maryland Clark Lake telescope to obtain a detailed map of
the Coma cluster at 30.9\,MHz. She then derived the total diffuse flux after 
subtraction of the contributions of discrete sources.

A substantial number of workers have measured this flux at frequencies above 
30.9\,MHz. (see, for example, the summaries in Kim et al., 1990; Giovannini et
al., 1993, and Deiss et al., 1997). At high frequencies, measurements at 2.7 
and 4.85\,GHz fall below a simple extrapolation of the spectral index derived 
from data at lower frequencies. Schlickeiser et al. (1987) have suggested that
this may indicate a break in the spectral index at higher frequencies, and 
Deiss et al. (1997) provide evidence that this is an instrumental effect. 
In either case, we believe it is most reasonable to use only the measurements 
at or below 1.4\,GHz to derive a spectral index for the cluster.

The data between 30.9\,MHz and 1.4\,GHz are well fitted with $\alpha = 1.16$. 
We will use this value hereafter, although we note that many authors,
and specifically Hwang (1997), and En{\ss}lin and Biermann (1998) use the 
spectral index derived from the entire dataset which is best fit with a 
spectral index of 1.36.

We turn now to the question of the magnetic field in the cluster. A number of 
estimates of this field have been made. The simplest approach is to assume 
equipartition between the energy in the magnetic field and the relativistic 
electrons. This approach has been used by Giovannini et al. (1993), who 
obtained a magnetic field of $0.5 \mu$G. 

Kim et al. (1990) measured the Faraday rotation of background radio sources 
shining through the cluster with background sources outside of the cluster
region. This provided an estimate of the excess rotation through the cluster,
which is due to free electrons produced by the thermal X-ray gas in the 
cluster. These workers obtained a value of $1.7 \pm 0.9 \mu$G in the 
cluster. Feretti et al. (1995) measured the de-polarization of the radio 
emission from an active galaxy near the core of the cluster and obtained a 
magnetic field of $6 \pm 1 \mu$G. Each of these methods has its own set 
of difficulties; the existing data and theoretical arguments suggest the 
field is between 0.1 and $7 \mu$G.

\begin{figure}
\plotone{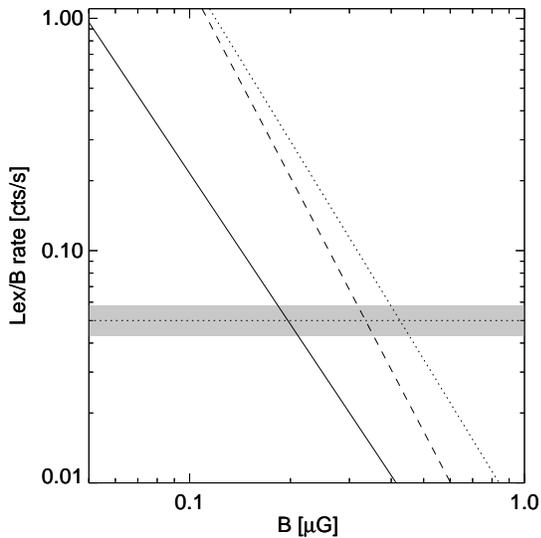}
\caption{The EUVE count rate as a function of magnetic field in the
cluster for a number of synchrotron emission spectral indices. The solid line 
corresponds to $\alpha$ = 1.16. The dotted line corresponds to $\alpha$ = 1.36.
The dashed line corresponds to a power law radio component with 10\,Jy at 
30.9\,MHz and $\alpha$=1.75 (see text). The shaded region shows 
the total cluster EUV excess with one sigma errors. \label{rateb}}
\end{figure}

Calculations of the inverse Compton emission from gas in clusters of galaxies 
typically assume a spherical source and a spatially constant spectral index 
for the electrons as derived from the integrated flux density of the
synchrotron emission at different frequencies. Using Equation 4 and a model of
this type, we have calculated the expected count rates in the  Lexan/Boron
filter of the Deep Survey Telescope of the Extreme Ultraviolet Explorer
(Bowyer and Malina, 1991) as a function of magnetic field in the cluster for a
spectral index of $\alpha = 1.16$. The results are shown in Figure\
\ref{rateb} as a solid line. In this Figure we also show as a dotted
horizontal line the excess EUV count rate above that expected from the
X-ray-emitting cluster gas as obtained by Lieu et al. (1996b); the shaded
region shows the one sigma statistical errors in the measurement. These data
suggest that the magnetic field in the cluster is $B \simeq 0.2 \mu$G. A
spectral index $\alpha = 1.36$ (which we have argued above is inappropriate)
is shown as a dotted line. This would suggest $B \simeq 0.4 \mu$G.

Combinations of lower magnetic fields and steeper power law indices produce 
inverse Compton flux above the actual detected EUV flux. A straightforward 
conclusion is that the EUV data limit these parameters to the values indicated.
However, an alternative possibility is that a mechanism is present which 
reduces the inverse Compton flux to the level actually observed.

The only mechanism we have identified which could reduce the EUV flux from 
this cluster is absorption by an extended neutral hydrogen halo. Such a halo 
would be extremely difficult to detect by its 21-centimeter emission. However,
M. Urry (private communication) has obtained Space Telescope spectra of the 
QSO 1258+285, which is in the line of sight of the cluster. This spectrum 
includes the red-shifted wavelengths corresponding to Lyman-$\alpha$ 
absorption from neutral hydrogen in the cluster. The absorption observed is 
unsaturated and shows that N(H) $\lesssim 3 \times 10^{13}$, which is
clearly insufficient to absorb any appreciable EUV radiation.

\section{The Spatial Distribution of the Inverse Compton Emission}

\begin{figure}
\plotone{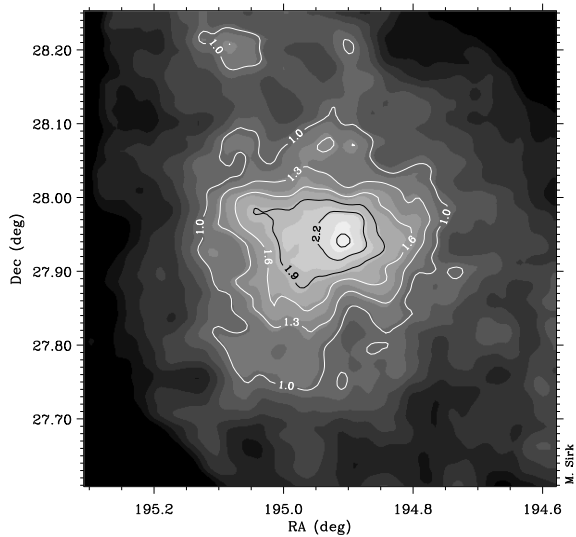}
\caption{The spatial distribution of the EUV emission from the
Coma cluster in excess of that expected from that produced by the
high temperature X-ray emitting gas.
The contour levels are counts s$^{-1}$ deg$^{-2}$.
The image was created by convolving the raw data with a Gaussian
with a FWHP equal to the on-axis point spread function of the Deep Survey
telescope ($37 \arcsec$),and
then subtracting a mean background estimated far from the cluster
center. \label{euvima}}
\end{figure}

The EUVE data provide more information than simply the total EUV flux from the
Coma cluster; they also provide information on the spatial distribution of this
flux. In Figure\ \ref{euvima} we show an isophotal map of this flux.
The emission is clearly not spherically distributed but is elongated in the 
east-west direction. We note that a fraction of the EUV flux is produced
by the high temperature X-ray emitting plasma. This portion of the flux is 
almost spherical (White, et al, 1993) and hence the asymmetric distribution of 
the total flux is the result of whatever process is producing the excess EUV 
flux.

We have quantified the distribution of the excess EUV flux. We first subtracted
the EUV flux produced by the x-ray emitting gas with T=8.2\,keV (Hatsukade, 
1990), a central electron density of 2.89$\times$10$^{-3}$\,cm$^{-3}$ and the 
best fit King profile provided by Briel et al (1992). We then fit a two 
dimensional Gaussian to the excess. The fit is reasonable (red. $\chi^2 = 1.4$
for $40 \times 40$ pixels of $1\arcmin \times 1\arcmin$ size).
The excess flux has a FWHP of $19.3\arcmin \times 12.6 \pm 1.5\arcmin$.

A number of studies of the spatial distribution of the Coma radio halo at
different frequencies have been carried out. These all show an east-west
elongation. In Table 1 we provide results from some of the more detailed radio
investigations in which the authors have fit a two dimensional Gaussian to
their data. As stated previously, the X-ray emission is almost spherical and 
previous workers have not cited parameters for an asymmetric profile. For this
work,we have quantified the X-ray distribution by fitting a two dimensional 
King profile to the X-ray map of White, et al (1993). The FWHP of the X-ray 
emission derived from this fit is 17$\arcmin$ by 15$\arcmin$.
\begin{table*}[h]
  \caption{Coma Halo Shapes}
  \begin{tabular}{lcccl}
   \hline\hline 
& size & aspect & & \\
& (FWHP) & ratio & $\nu$ & reference\\
    \hline
Radio  & $31 \times 15$ & 2.1 & 30.9\,MHz & Henning (1989) \\
 & $28 \times 20$ & 1.4 & 326\,MHz & Giovannini et al. (1993) \\
 & $19 \times 14$ & 1.4 & 1.4\,GHz & Kim et al. (1990) \\
X-ray & $17 \times 15$ & 1.1 & & White et al. (1993)\\
EUV  & $19.3 \times 12.6$ & 1.5 & & This work\\
    \hline
  \end{tabular}
\end{table*}

In Table 1 we also show an ``aspect ratio'' for the results obtained.
This parameter is defined as the FWHP size of the larger dimension divided by
the FWHP of the smaller dimension.
The asymmetry of the EUV emission more nearly resembles the radio emission
than the gravitationally bound, X-ray thermal gas emission. This is compelling
evidence that the production mechanism for the EUV flux in the Coma cluster
is related to  non-thermal processes  and is not the result of a
gravitationally bound thermal gas.

\begin{figure}
\plotone{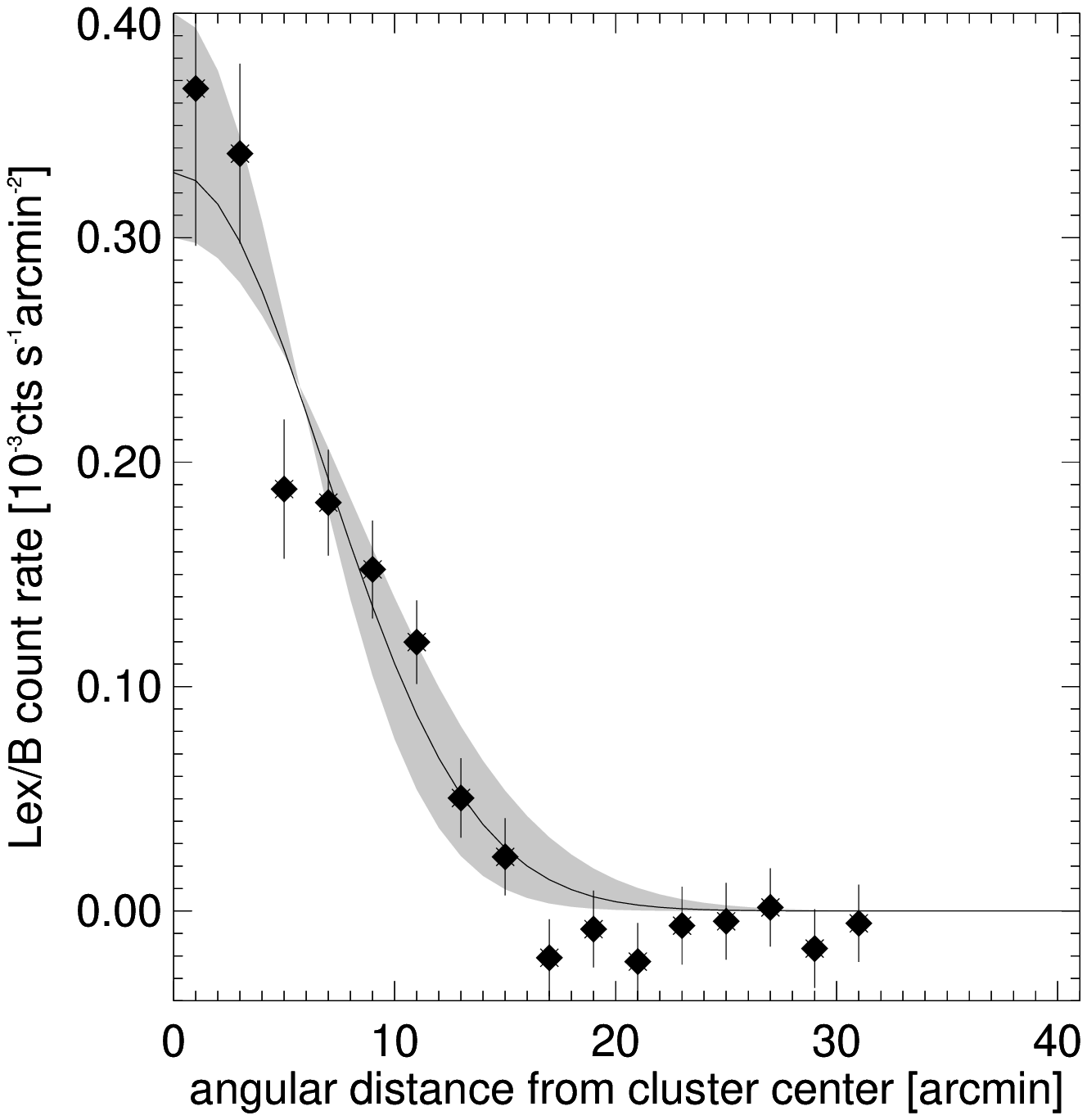}
\caption{The azimuthally averaged radial intensity profile of the 
cluster's EUV excess. The EUV radiation is best fit by a Gaussian distribution
of FWHP$ = 15.8\arcmin$; the distribution is shown as a solid line, the one 
sigma error of this fit is indicated by the gray shaded region. \label{rad2}}
\end{figure}

In Figure\ \ref{rad2} we show the azimuthally averaged excess EUV count rate as
a function of distance from the cluster center. The data are well fit by a
Gaussian with a size of FWHP$ = 15.8\arcmin$ (red. $\chi^2 = 1.16$) which is
shown as a solid line. The one sigma error of the fit is shown as a gray 
shaded region. Within the errors the EUV excess emission shows the same spatial
azimuthally averaged distribution as the radio emission at 1.4\,GHz 
(FWHP$ = 15.2\arcmin$).

Several groups (Giovannini et al., 1993, Deiss et al., 1997) have shown that
at higher frequencies the spectral index of the synchrotron emission near the
core of the cluster is substantially flatter than  
that further out. Deiss et al. (1997) and Deiss (1997) have developed 
formalisms which relate the distribution of the spectral indices to the 
surface brightness distribution.  In addition, they have developed a 
semi-empirical relationship between the radio halo size and spectral index 
distribution at two different frequencies. They show that the variation of the
spectral index is interrelated with the increasing cluster size at lower 
frequencies. This follows from the requirement that a larger halo size at 
lower frequencies requires a flux redistribution from the cluster center to 
larger radii. This effect is observed for the Coma cluster in the frequency 
range 326\,MHz -- 1.4\,GHz

However, at frequencies $\le 326$\,MHz the radio halo size is essentially
independent of frequency; the average FWHP at 30.9\,MHz ($\sim 23\arcmin$) is
about the same as observed at 326\,MHz ($\sim 24\arcmin$). Following Deiss et
al. (1997) the $\alpha$ distribution will be constant. The existing data 
therefore imply a spectral index distribution at low
frequencies that is almost constant throughout the cluster. Further, a
decreasing halo size at lower frequencies would require a spectral index
distribution that is steeper near the cluster center than at larger radii.

The radio halo size in the frequency range 30.9 -- 326\,MHz is significantly
larger than the size of the EUV excess emission. This has far-reaching
consequences for the interpretation of the Coma EUV excess. As can be seen
from eq.\ (4) the magnetic field distribution in the intracluster medium is
related to the inverse Compton and synchrotron emission by:

\begin{equation}
B^{-(\alpha + 1)}(r) \propto \frac{F_{ic}(r)}{F_{sync}(r)} \hspace{10pt}.
\end{equation}

\begin{figure}
\plotone{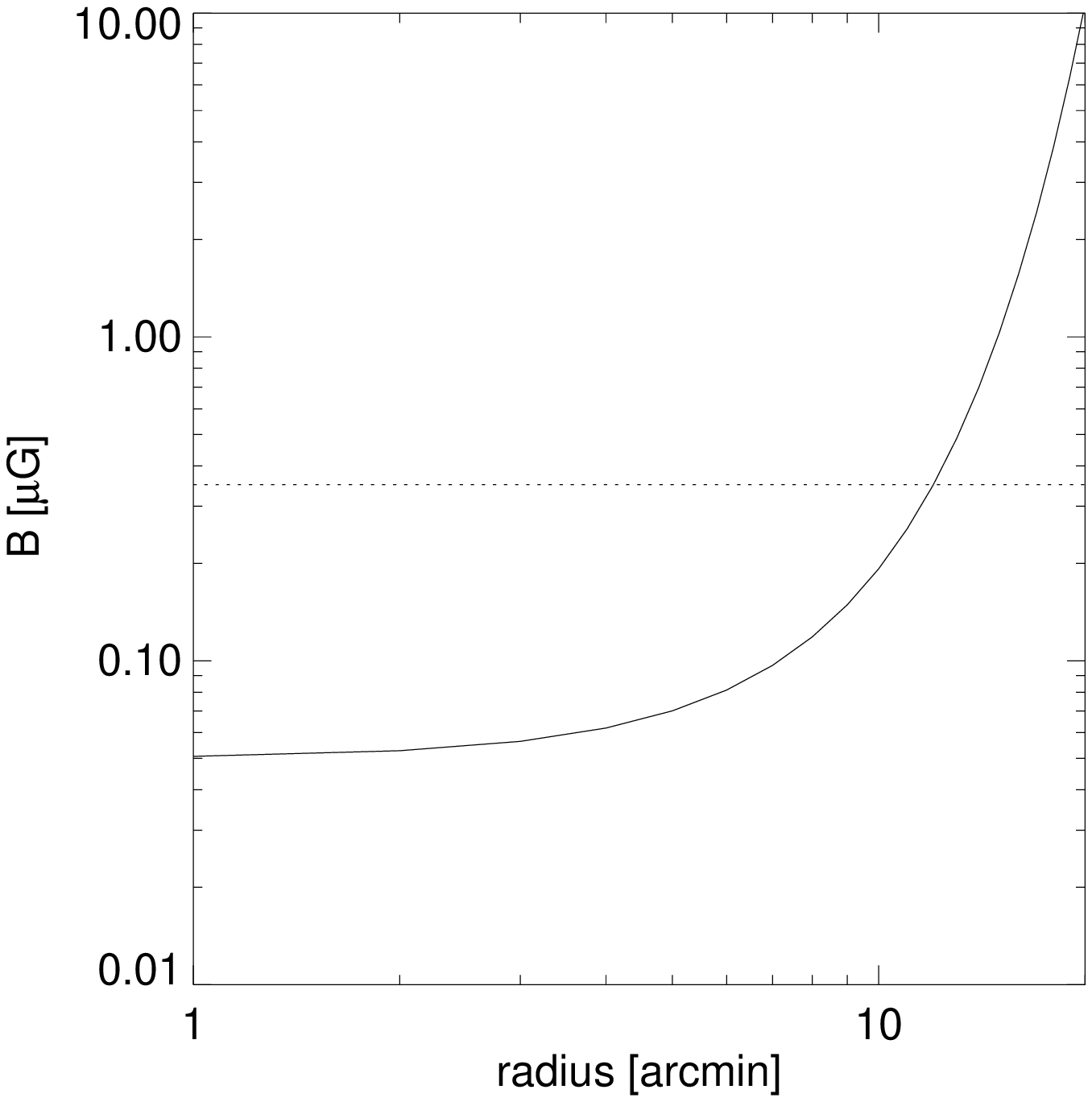}
\caption{The radial distribution of the magnetic field as derived
from the radial EUV emission profile and the radio profile in the frequency
range 30.9 -- 326\,MHz. The solid line shows the field required if the
EUV emission is produced by an extrapolation of electrons producing the 
observed radio emission. The dotted line shows the field required for the 
additional component of electrons we propose herein. \label{bfield}}
\end{figure}

If we assume that the spectral index distribution is radius independent and
use the fact that $F_{ic}$ and $F_{sync}$ have halo sizes of
FWHP$_{ic} = 15.8\arcmin$ and FWHP$_{sync} = 24\arcmin$, the magnetic field
strength distribution $B(r)$ is defined by the ratio $F_{ic}/F_{sync}$ and
must increase towards larger radii. Figure\ \ref{bfield} shows the radial
distribution of the magnetic field strength for this case. However, a radially
increasing magnetic field appears to be unphysical since this field must
soon meld with the cluster field at larger radii.

We conclude that there is no straightforward way to produce the EUV halo with
an electron population which is an extrapolation of the observable synchrotron
producing electrons.

In order to explain the EUV halo as inverse Compton emission from a
population of electrons with reasonable properties, we must postulate an
additional population of low energy relativistic electrons in the central part
of the cluster. These electrons will, by necessity, have an asymmetric 
distribution defined by the EUV emission profile but for simplicity, we use 
the azimuthally averaged Gaussian profile in the following. 
R. Lieu (to be published) has made a power-law fit to the excess EUV flux 
using data from the somewhat different bandpasses of EUVE and the ROSAT PSPC. 
He obtained a spectrum with a photon index of $\approx 1.75$. If our scenario 
is correct, the spectral index of the EUV emission and the electron 
distribution will be the same, hence we use this spectral index for our 
additional population of electrons. The only other observational constraint is
imposed by the fact that these electrons cannot produce observable
synchrotron emission. Accordingly, we use the one sigma error in the 30.9\,MHz
observation (10\,Jy) as an upper limit for the radio emission of this
component.

The EUVE count rate for  this population of electrons is shown as a function of
magnetic field strength in Figure\ \ref{rateb}.
The magnetic field that is required if the total EUV excess flux is produced
by this population of relativistic electrons is $\sim 0.35\mu$G. This field is
constant throughout the EUV emitting region; we show this as a dotted line in 
Figure\ \ref{bfield}.
Halo sizes with FWHP$_{sync}$ less than this value lead to decreasing values
of $B(r)$ towards larger radii. A decreasing radio halo size at lower
frequencies also affects the spectral index distribution across the cluster,
producing a spectral index distribution which decreases with cluster radius.
This results in a somewhat flatter, but nonetheless decreasing, magnetic
field distribution across the cluster.

\begin{figure}
\plotone{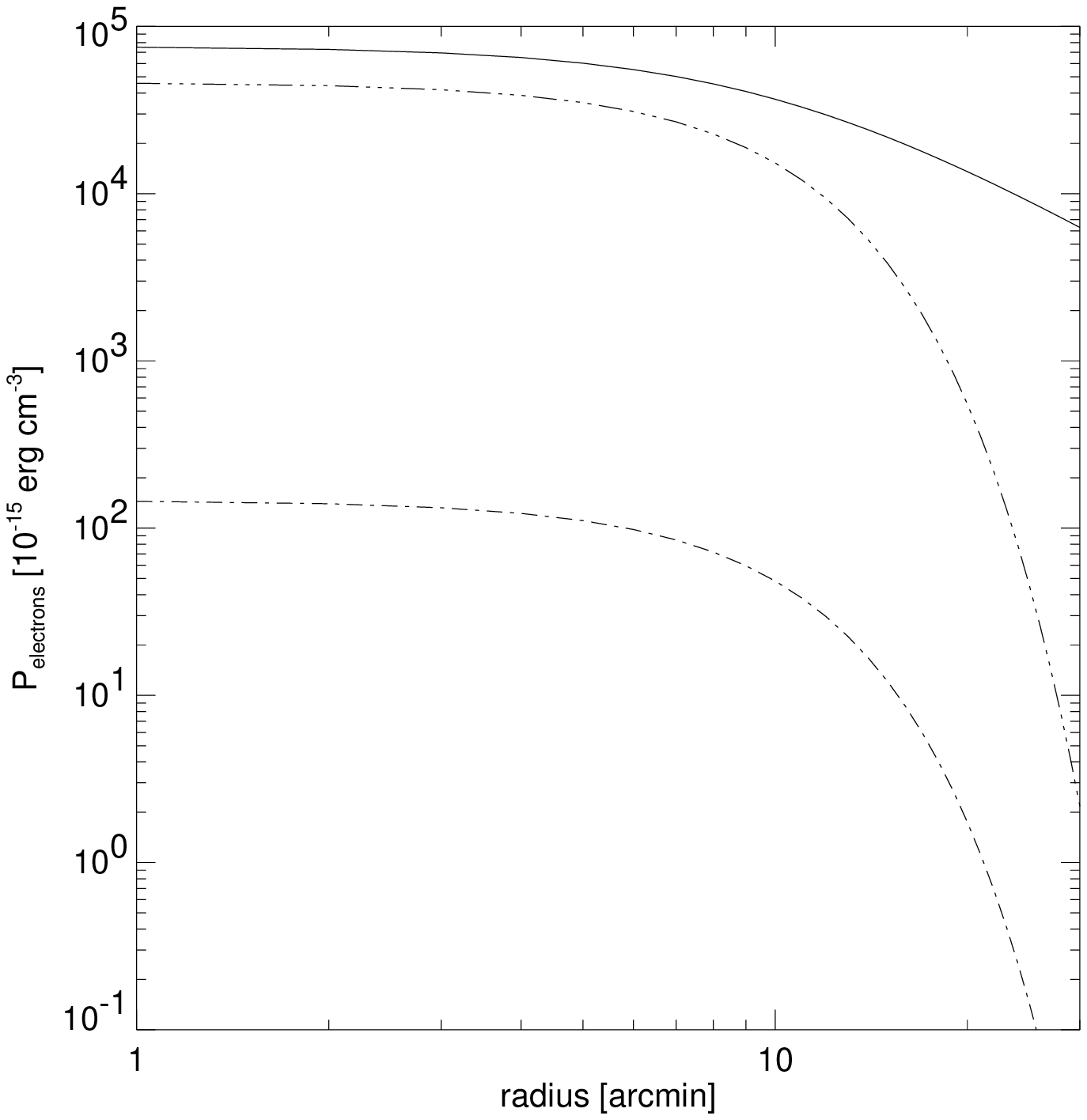}
\caption{The pressure of the cosmic ray electrons as a function of
radius in comparison with the pressure of the X-ray emitting gas (solid line).
The dash-dotted line shows the pressure for a population of relativistic 
electrons with a power law spectrum of $\alpha = 1.75$, an intensity of
10\,Jy at 30.9\,MHz, and a low energy cutoff at 0.16\,GeV. With a ten times 
lower cutoff energy, the pressure distribution for this population is given by
the dash-dot-dot-dotted line. \label{press}}
\end{figure}

We have computed the pressure of the additional relativistic electron 
population we have introduced to explain the EUV flux profile. 
The pressure of the relativistic electrons is 1/3 of the energy density of 
these electrons; as a first order approximation we used a spatial distribution
with a FWHP of 15.8$\arcmin$ which is the average value obtained from the EUV 
emission profile assuming a spherically symmetric Gaussian fit to the EUV 
surface brightness profile. We have normalized the result to the total energy,
$E_{tot}$, of the relativistic electrons. $E_{tot}$ can be obtained by an 
integration of the electron energy spectrum given by eq. (1) which leads to:
\begin{equation}
E_{tot} = \int_{E_{low}}^{\infty}N_{0}E^{-\gamma} E dE = \frac{1}{\gamma-2}N_{0}E_{low}^{2-\gamma}.
\end{equation}
The pressure of the relativistic electrons depends upon the spectral index of
the electrons and the low energy cutoff of these electrons which is not known.
These electrons must extend to energies as low as 0.16\,GeV in order to explain
the EUV emission by the inverse Compton mechanism and we use this cutoff.
Although this is certainly physically unrealistic, it will provide a minimum
pressure estimate. An electron energy cutoff at 0.016\,GeV results in a higher
 pressure but this is still well below that of the X-ray gas. The total energy
of these electrons, assuming $E_{low} = 0.16$\,GeV, is 
$5.5 \times 10^{60}$\,erg. For $E_{low} = 0.016$\,GeV we compute a total 
energy of $1.2 \times 10^{63}$\,erg. These results are shown in Figure\ 
\ref{press}. 

We have computed the pressure for the X-ray emitting gas (electrons and ions) 
as a function of cluster radius for a cluster temperature of $kT = 8.2$\,keV 
(Hatsukade 1990), a central electron density of 2.89$\times$10$^{-3}$cm$^{-3}$,
and the best fit King profile provided by Briel, Henry, and B\"ohringer (1992)
and display these results in Figure\ \ref{press} as a dotted line.

\section{Conclusions}

We have examined the hypothesis that the EUV radiation from the Coma cluster
is due to inverse Compton scattering of low energy cosmic ray electrons against
the 3$^{\circ}$\,K black-body background radiation. The total integrated EUV
emission produced by cosmic ray electrons which are a low energy extrapolation
of higher energy electrons, known to be present from their synchrotron 
emission, gives results which are consistent with the range of estimates of 
the magnetic field in the cluster.

We next consider the two dimensional spatial distribution of the EUV emission.
This emission does not follow the distribution of the gravitationally bound
X-ray gas, but rather exhibits an asymmetric distribution similar to that
exhibited by the radio emission. This suggests a non-thermal origin for the
EUV emission rather than a gravitationally constrained thermal gas.

We show from a comparison of the size of the EUV halo and the radio halo
that the EUV emission cannot be produced by inverse Compton radiation from
electrons which are an extrapolation of the distribution which
produces the observed radio emission. We develop a model for the EUV emission
which is self-consistent and fits the existing data. This model requires an
additional component of low energy cosmic rays.

Inverse Compton EUV emission is surely present at some level in clusters of
galaxies with radio halos. However, it may well be masked by emission from
some other more dominant source mechanism. A test of the inverse Compton
scattering hypothesis as the source of the EUV flux in the Coma cluster would
be provided by a measurement of the size of the radio halo at $\sim 1$\,MHz.
Unfortunately, this is a challenging measurement because of instrumental
limitations and ionospheric effects. In addition, at these low frequencies 
self-absorption could affect the surface brightness profile by reducing the 
flux near the cluster center while increasing the halo size; this effect would 
have to carefully be considered when interpreting such a measurement.

\acknowledgments
{\bf Acknowledgements}\\
We thank Bruno Deiss, Philipp Kronberg, Hans B\"ohringer, Thorsten En{\ss}lin,
Peter Biermann, Michael Lampton, Greg Sarazin, and Chorng-Yuan Hwang for
useful discussions and suggestions. We thank Martin Sirk for the isophotal map
of the EUV emission.  T.W.B. acknowledges the support from the 
Alexander-von-Humboldt-Stiftung (AvH) by a Feodor-Lynen Fellowship. This work 
has been supported by NASA contract NAS 5-30180.

\end{document}